\documentclass[12pt]{article}
\usepackage{epsf}
\title{ {\bf The $\mu\rightarrow e\gamma$ and $\tau\rightarrow \mu\gamma$
decays in the general two Higgs Doublet model with the inclusion
of one universal extra dimension.}}
\author{\vspace{1cm}\\
        {\bf E. O. Iltan}
        \thanks{E-mail address:
        eiltan@heraklit.physics.metu.edu.tr}
 \\
        Physics Department, Middle East Technical University \\
        Ankara, Turkey\\}

\date{}

\begin{document}
\setlength{\baselineskip}{24pt}
\maketitle
\setlength{\baselineskip}{7mm}
\begin{abstract}
We study the effect of one universal extra dimension on the
branching ratios of the lepton flavor violating  processes
$\mu\rightarrow e\gamma$ and $\tau\rightarrow \mu\gamma$ in the
general two Higgs doublet model. We observe that these new effects
are tiny for the small values of the compactification radius $R$.
Furthermore,  we see that these effects are comparable with the
branching ratio obtained without including extra dimension, if the
neutral Higgs bosons are nearly degenerate and the complexity of
the Yukawa coupling, inducing the vertex tau-tau-$h^0\,(A^0)$, is
large.
\end{abstract}
\thispagestyle{empty}
\newpage
\setcounter{page}{1}
\section{Introduction}
Lepton Flavor Violating (LFV) interactions are rich from the
theoretical point of view since they exist at the loop level and
the related measurable quantities contain number of free
parameters of the model used. In such decays, the assumption of
the non-existence of Cabibbo-Kobayashi-Maskawa (CKM) type matrix
in the leptonic sector, in the framework of the standard model
(SM), forces one to search the physics beyond. The improvement of
the experimental measurements of the LFV processes make it
possible to understand the new physics effects more accurately.
One of the candidate model beyond the SM is the general two Higgs
doublet model (2HDM), so called the model III. In this model the
LFV interactions are induced by the internal neutral Higgs bosons
$h^0$ and $A^0$ and the Yukawa couplings, appearing as free
parameters, can be determined by the experimental data. The
$\mu\rightarrow e\gamma$ and $\tau\rightarrow \mu\gamma$ are the
examples of LFV interactions and the current limits for their
branching ratios ($BR$'s) are $1.2\times 10^{-11}$ \cite{Brooks}
and $1.1\times 10^{-6}$ \cite{Ahmed} respectively. In the
literature, there are several studies on the LFV interactions in
various models. Such interactions are investigated in a model
independent way in \cite{Chang}, in supersymmetric models
\cite{Barbieri1}, in the framework of the 2HDM \cite{Iltan1,
Diaz}.

In this work, we study the LFV processes $\mu\rightarrow e\gamma$
and $\tau\rightarrow \mu\gamma$ in the framework of the model III,
with the addition of one extra spatial dimension. Higher
dimensional scenarios has been induced by the string theories as a
possible solution to the hierarchy problem of the standard model
(SM) and there is an extensive work on this subject in the
literature \cite{Arkani}-\cite{Lam}. The idea is that the ordinary
four dimensional SM is the low energy effective theory of a more
fundamental one lying in larger dimensions and the extra
dimensions over the 4-dimension are compactified on a circle of a
radius $R$, which is a typical size of an extra dimension. This
size has been studied using the present experimental data in
several works \cite{Antoniadis2,Carone} and estimated as large as
few hundereds of GeV \cite{Arkani, Antoniadis1,Antoniadis3}, not
to contradict with the experiments. Furthermore, the loop effects
induced by the internal top quark are sensitive to the KK mode
contributions and the size of the extra dimensions has been
estimated in the range $200-500\,  GeV$, using electroweak
precision measurements \cite{Appelquist}, the \( B-\bar{B} \)
-mixing \cite{Papavassiliou},\cite{Chakraverty} and the flavor
changing process $b \to s \gamma$ \cite{Agashe}.

In the 4 dimensions, the extra dimension takes the form of
Kaluza-Klein (KK) excitations of fields, 2HDM fields in our case,
with masses $\sim n / R$. In the case that all 2HDM fields lie in
the extra dimension, the extra dimensional momentum is conserved
and, in our case, the coupling of Higgs boson-lepton-lepton vertex
involves one external zero mode lepton and internal KK mode of
lepton and neutral Higgs boson. Such extra dimensions are called
universal extra dimensions (UED).

In the LFV $\mu\rightarrow e\gamma$ and $\tau\rightarrow
\mu\gamma$ decays, the effect of one UED is carried by the KK mode
of internal neutral Higgs fields, $h^0$, $A^0$, and internal
lepton fields, at one loop order, in the model III version of the
2HDM. The non-zero KK modes of neutral Higgs fields $H$ have
masses $\sqrt{m_{H}^2+m_n^2}$ with $m_n=n/R$. Here $m_n=n/R$ is
the mass of $n$'th level KK particle where R is the
compactification radius. Similarly, the non-zero KK modes of
lepton doublets (singlets) have the left (right) handed lepton
fields with masses $\sqrt{m_{l_i}^2+m_n^2}$ and the right (left)
handed ones with masses $m_n$. It should be noted that the KK
spectrum at each excitation level is nearly degenerate for
different lepton flavors since their masses are smaller compared
to the mass $m_n=n/R$.

This work is devoted to the the effect of one UED on the BR of the
LFV processes $\mu\rightarrow e\gamma$ and $\tau\rightarrow
\mu\gamma$ in the framework of the model III. It is observed that
these new effects are tiny for the small values of the
compactification radius. Furthermore, these effects are comparable
with the BR obtained without including extra dimension, in the
case that the neutral Higgs bosons are nearly degenerate and the
complexity of the Yukawa coupling is large.

The paper is organized as follows: In Section 2, we present the
BR's of LFV interactions $\mu\rightarrow e\gamma$ and
$\tau\rightarrow \mu\gamma$ in the model III version of the 2HDM
with the inclusion of one universal extra dimension. Section 3 is
devoted to discussion and our conclusions.
\section{The LFV interactions $\mu\rightarrow e\gamma$ and $\tau\rightarrow
\mu\gamma$ in the general two Higgs Doublet model with the
inclusion of one universal extra dimension.}
In the model III version of the 2HDM the flavor changing neutral
currents (FCNC) at tree level is permitted and the LFV
interactions exist with larger BR`s, compared to ones obtained in
the SM. The main parameters in this calculation are the new Yukawa
couplings, that can be chosen complex. The addition of one spatial
UED brings new contributions to the BR`s of LFV processes at one
loop order and the part of the Lagrangian responsible for these
interactions is the Yukawa Lagrangian, which reads in 5 dimension
\begin{eqnarray}
{\cal{L}}_{Y}=
\eta^{E}_{5\, ij} \bar{l}_{i} \phi_{1} E_{j}+
\xi^{E}_{5\, ij} \bar{l}_{i} \phi_{2} E_{j} + h.c. \,\,\, ,
\label{lagrangian}
\end{eqnarray}
with $5$-dimensional Yukawa couplings $\eta^{E}_{5\, ij}$,
$\xi^{E}_{5\, ij}$, where $i,j$ are family indices of leptons,
$\phi_{i}$ for $i=1,2$, are the two scalar doublets, $l_{i}$ and
$E_{j}$ are lepton doublets and singlets respectively. These
fields are the functions of $x^\mu$ and $y$, where y is the
coordinate represents the $5$`th dimension. The Yukawa couplings
$\eta^{E}_{5 \,ij}$, $\xi^{E}_{5\, ij}$ are dimensionful and
rescaled to the ones in 4-dimension as $\eta^{E}_{5\, ij}=\sqrt{2
\pi R}\, \eta^{E}_{ij}$, and  $\xi^{E}_{5\, ij}=\sqrt{2 \pi
R}\,\xi^{E}_{5 ij}$. Here $\phi_{1}$ and $\phi_{2}$ are chosen as
\begin{eqnarray}
\phi_{1}=\frac{1}{\sqrt{2}}\left[\left(\begin{array}{c c}
0\\v+H^{0}\end{array}\right)\; + \left(\begin{array}{c c}
\sqrt{2} \chi^{+}\\ i \chi^{0}\end{array}\right) \right]\, ;
\phi_{2}=\frac{1}{\sqrt{2}}\left(\begin{array}{c c}
\sqrt{2} H^{+}\\ H_1+i H_2 \end{array}\right) \,\, ,
\label{choice}
\end{eqnarray}
and the vacuum expectation values are
\begin{eqnarray}
<\phi_{1}>=\frac{1}{\sqrt{2}}\left(\begin{array}{c c}
0\\v\end{array}\right) \,  \, ;
<\phi_{2}>=0 \,\, .
\label{choice2}
\end{eqnarray}
With this choice, the SM particles can be collected in the first doublet
and the new particles in the second one. The part which produce FCNC at tree
level is
\begin{eqnarray}
{\cal{L}}_{Y,FC}=
\xi^{E}_{5\, ij} \bar{l}_{i} \phi_{2} E_{j} + h.c. \,\, .
\label{lagrangianFC}
\end{eqnarray}
Here the Yukawa matrices $\xi^{E}_{5\, ij}$ have in general
complex entries and they are the source of CP violation. Note that
in the following we replace $\xi^{E}$ with $\xi^{E}_{N}$ where "N"
denotes the word "neutral". The five dimensional lepton doublets
and singlets have both chiralities and the 4-dimensional
Lagrangian is constructed by expanding these  fields into their KK
modes. Besides, the extra dimension denoted by $y$ is compactified
on a circle of radius $R$ and the zero modes of the wrong
chirality ($l_{i R}$, $E_{i L}$) are projected out by
compactification of the fifth dimension $y$  on an $S^1/Z_2$
orbifold, namely $Z_2:y \to -y$. The KK decompositions of the
lepton and Higgs fields read
\begin{eqnarray}
\phi_{1,2}(x,y ) & = & {1 \over {\sqrt{2 \pi R}}} \left\{
\phi_{1,2}^{(0)}(x) + \sqrt{2}
\sum_{n=1}^{\infty}  \phi_{1,2}^{(n)}(x) \cos(ny/R)\right\}\nonumber\\
l_i (x,y )& = & {1 \over {\sqrt{2 \pi R}}} \left\{ l_{i
L}^{(0)}(x) + \sqrt{2} \sum_{n=1}^{\infty}  \left[l_{i L}^{(n)}(x)
 \cos(ny/R) + l_{i R}^{(n)}(x) \sin(ny/R)\right]\right\}\nonumber\\
E_{i}(x,y )& = & {1 \over {\sqrt{2 \pi R}}} \left\{ E_{i
R}^{(0)}(x) + \sqrt{2} \sum_{n=1}^{\infty}  \left[E_{i R}^{(n)}(x)
\cos(ny/R) + E_{i L}^{(n)}(x) \sin(ny/R)\right]\right\} \, ,
\end{eqnarray}
where $\phi_{1,2}^{(0)}(x)$, $l_{i L}^{(0)}(x)$ and $E_{i
R}^{(0)}(x)$ are the 4-dimensional Higgs doublets, lepton doublets
and lepton singlets respectively. Here $L$ and $R$ denote chiral
projections $L(R)=1/2(1\mp \gamma_5)$ and they are four
dimensional. Each non-zero KK mode of Higgs doublet $\phi_{2}$
($\phi_{1}$) includes a charged Higgs of mass
$\sqrt{m_{H^\pm}^2+m_n^2}$ ($m_n$), a neutral CP even Higgs of
mass $\sqrt{m_{h^0}^2+m_n^2}$ ($\sqrt{m_{H^0}^2+m_n^2}$), a
neutral CP odd Higgs of mass $\sqrt{m_{A^0}^2+m_n^2}$ ($m_n$) with
$m_n=n/R$. Here $m_n=n/R$ is the mass of $n$'th level KK particle
where R is the compactification radius. Similarly, the non-zero KK
modes of lepton doublets (singlets) have the left (right) handed
lepton fields with masses $\sqrt{m_{l_i}^2+m_n^2}$ and the right
(left) handed ones with masses $m_n$. Notice that the KK spectrum
at each excitation level is nearly degenerate for different lepton
flavors since their masses are smaller compared to the mass
$m_n=n/R$.

Now we will consider the lepton flavor violating processes
$\mu\rightarrow e\gamma$ and $\tau\rightarrow \mu\gamma$ with the
addition of one spatial dimension. These processes are good
candidates for searching the new physics beyond the SM and for the
determination of the free parameters existing. Here we take into
account only the neutral Higgs contributions in the lepton sector
of the model III and, therefore, the neutral Higgs bosons $h^0$
and $A^0$ are responsible for these interactions. The addition of
one extra spatial dimension brings new contribution due to the
internal $h^{0 n}$- $l_i^n \,(l_{1,2,3}=\tau, \mu, e)$, $A^{0 n}$-
$l_i^n$ KK modes and these contributions are calculated by taking
the vertices involving one zero mode and two non-zero modes (see
Fig. \ref{fig1}). In the calculations, the on-shell
renormalization scheme is used. In this scheme, the self energy
diagrams for on-shell leptons vanish since they can be written as
\begin{eqnarray}
\sum(p)=(\hat{p}-m_{l_1})\bar{\sum}(p) (\hat{p}-m_{l_2})\,\, ,
\label{self}
\end{eqnarray}
However, the vertex diagram $a$ and $b$ in Fig. \ref{fig1} gives
non-zero contribution. Taking only $\tau$ lepton for the internal
line, the decay width $\Gamma$ reads as
\begin{eqnarray}
\Gamma (\mu\rightarrow e\gamma)=c_1(|A_1|^2+|A_2|^2)\,\, ,
\label{DWmuegam}
\end{eqnarray}
where
\begin{eqnarray}
A_1&=&Q_{\tau} \frac{1}{48\,m_{\tau}^2} \Bigg (6\,m_\tau\,
\bar{\xi}^{E *}_{N,\tau e}\, \bar{\xi}^{E *}_{N,\tau \mu}\, \Big(
F (z_{h^0})-F (z_{A^0})\Big ) \nonumber \\ &+&
m_{\mu}\,\bar{\xi}^{E *}_{N,\tau e}\, \bar{\xi}^{E}_{N,\tau \mu}\,
\Big(G (z_{h^0})+G (z_{A^0})+\sum_{n=1}^{\infty}
\frac{m_\tau^2}{(m^{extr}_{\tau})^2} (G (z_{n, h^0})+G (z_{n,
A^0}))\Big) \Bigg)
\nonumber \,\, , \\
A_2&=&Q_{\tau} \frac{1}{48\,m_{\tau}^2} \Bigg (6\,m_\tau\,
\bar{\xi}^{D}_{N,\tau e}\, \bar{\xi}^{D}_{N,\tau \mu}\, \Big(F
(z_{h^0})-F (z_{A^0})\Big)\nonumber \\ &+&
m_{\mu}\,\bar{\xi}^{D}_{N,\tau e}\, \bar{\xi}^{D *}_{N,\tau \mu}\,
\Big( G (z_{h^0})+G (z_{A^0})+ \sum_{n=1}^{\infty}
\frac{m_\tau^2}{(m^{extr}_{\tau})^2}(G (z_{n, h^0})+ G (z_{n,
A^0})) \Big) \Bigg)
 \,\, , \label{A1A2}
\end{eqnarray}
$c_1=\frac{G_F^2 \alpha_{em} m^3_{\mu}}{32 \pi^4}$. Here the
amplitudes $A_1$ and $A_2$ have right and left chirality
respectively. The decay width of the other LFV process
$\tau\rightarrow \mu\gamma$ can be calculated using the same
procedure and reads as
\begin{eqnarray}
\Gamma (\tau\rightarrow \mu\gamma)=c_2(|B_1|^2+|B_2|^2)\,\, ,
\end{eqnarray}
\label{DWtaumugam}
where
\begin{eqnarray}
B_1&=&Q_{\tau} \frac{1}{48\,m_\tau^2} \Bigg (6\,m_{\tau}\,
\bar{\xi}^{E *}_{N,\tau \mu}\, \bar{\xi}^{E *}_{N,\tau \tau}\,
\Big( F (z_{h^0})-F (z_{A^0}) \Big) \nonumber \\ &+&
m_{\tau}\,\bar{\xi}^{E *}_{N,\tau \mu}\, \bar{\xi}^{E}_{N,\tau
\tau}\, \Big( G (z_{h^0})+G (z_{A^0})+\sum_{n=1}^{\infty}
\frac{m_\tau^2}{(m^{extr}_{\tau})^2}(G (z_{n, h^0})+ G (z_{n,
A^0}))\Big)  \Bigg)
\nonumber \,\, , \\
B_2&=&Q_{\tau} \frac{1}{48\,m_\tau^2} \Bigg (6\,m_{\tau}\,
\bar{\xi}^{D}_{N,\tau \mu}\, \bar{\xi}^{D}_{N,\tau \tau}\, \Big(
F(z_{h^0})-F (z_{A^0})\Big)\nonumber \\ &+&
m_{\tau}\,\bar{\xi}^{D}_{N,\tau \mu}\, \bar{\xi}^{D *}_{N,\tau
\tau}\, \Big( G (z_{h^0})+G (z_{A^0})+\sum_{n=1}^{\infty}
\frac{m_\tau^2}{(m^{extr}_{\tau})^2}(G (z_{n, h^0})+G (z_{n,
A^0}))\Big)  \Bigg)
 \,\, , \label{B1B2}
\end{eqnarray}
and $c_2=\frac{G_F^2 \alpha_{em} m^5_{\tau}}{32 \pi^4}$. Here the
amplitudes $B_1$ and $B_2$ have right and left chirality,
respectively. The functions $F (w)$ and $G (w)$ in eqs.
(\ref{A1A2}) and (\ref{B1B2})are given by
\begin{eqnarray}
F (w)&=&\frac{w\,(3-4\,w+w^2+2\,ln\,w)}{(-1+w)^3} \, , \nonumber \\
G (w)&=&\frac{w\,(2+3\,w-6\,w^2+w^3+ 6\,w\,ln\,w)}{(-1+w)^4} \,\,
, \label{functions2}
\end{eqnarray}
where  $z_{H}=\frac{m^2_{\tau}}{m^2_{H}}$, $z_{n,
H}=\frac{m^2_{\tau}+(n/R)^2}{m^2_{H}+(n/R)^2}$,
$(m^{extr}_{\tau})^2=m^2_{\tau}+(n/R)^2$, $Q_{\tau}$ is the charge
of $\tau$ lepton. The Yukawa couplings $\bar{\xi}^{E}_{N,ij}$
appearing in the expressions are defined as
$\xi^{E}_{N,ij}=\sqrt{\frac{4\,G_F}{\sqrt{2}}}\,
\bar{\xi}^{E}_{N,ij}$. In our calculations  we take into account
only internal $\tau$-lepton contribution since, in our assumption,
the couplings $\bar{\xi}^{E}_{N, ij},\, i,j=e,\mu$, are small
compared to $\bar{\xi}^{E}_{N,\tau\, i}\, i=e,\mu,\tau$ due to the
possible proportionality of them to the masses of leptons under
consideration in the vertices and we used parametrization
\begin{eqnarray}
\bar{\xi}^{E}_{N,\tau l}=|\bar{\xi}^{E}_{N,\tau l}|\, e^{i
\theta_{l}} \,\, , \label{xi}
\end{eqnarray}
with the CP violating phase $\theta_{l}$ to extract the complexity
of these couplings. Furthermore, we take the Yukawa couplings for
the interactions lepton-KK mode of lepton-KK mode of Higgs bosons
($h^0$ and $A^0$) as the same as the ones for the interactions of
zero mode fields.
\section{Discussion}
The LFV $l_i\rightarrow l_j\gamma$ ($i\neq j$) interactions exist
at the loop level in the model III and the Yukawa couplings
$\bar{\xi}^D_{N,ij}, i,j=e, \mu, \tau$ are the essential
parameters used in the calculations of physical quantities related
to those decays. The Yukawa couplings are free parameters of the
theory and they can be fixed by present and forthcoming
experiments. In our calculations, we assume that the Yukawa
couplings $\bar{\xi}^{E}_{N,ij}$ is symmetric with respect to the
indices $i$ and $j$ and take $\bar{\xi}^{E}_{N,ij},\, i,j=e,\mu $,
as small compared to $\bar{\xi}^{E}_{N,\tau\, i}\, i=e,\mu,\tau$
since the strength of these couplings are related with the masses
of leptons denoted by the indices of them, similar to the
Cheng-Sher scenerio \cite{Sher}. For the $\mu\rightarrow e \gamma$
decay the Yukawa couplings $\bar{\xi}^{E}_{N,\tau \mu}$ and
$\bar{\xi}^{E}_{N,\tau e}$ play the main role. The first one is
restricted by using the experimental uncertainty, $10^{-9}$, in
the measurement of the muon anomalous magnetic moment and the
upper limit of $\bar{\xi}^{E}_{N,\tau \mu}$ is predicted as $30\,
GeV$ (see \cite{Iltananomuon} and references therein). For the
numerical values of the Yukawa coupling $\bar{\xi}^{E}_{N,\tau
e}$, we use the prediction $10^{-3}-10^{-2}\, GeV$ which respects
the experimental upper limit of BR of $\mu\rightarrow e \gamma$
decay, $BR\leq 1.2\times 10^{-11}$ and predicted value of
$\bar{\xi}^{E}_{N,\tau \mu}\leq 30\, GeV$ (see \cite{Iltan1} for
details). For the $\tau\rightarrow \mu \gamma$ decay the Yukawa
couplings $\bar{\xi}^{E}_{N,\tau \tau}$ and $\bar{\xi}^{E}_{N,\tau
\mu}$ play the main role and for $\bar{\xi}^{E}_{N,\tau \tau}$ we
use numerical values larger than the upper limit of
$\bar{\xi}^{E}_{N,\tau \mu}$.

The addition of one extra spatial dimension brings new
contribution to the BR of the decays under consideration and this
contribution emerges from the KK excitations of the lepton and
Higgs fields. In the case that all the fields live in higher
dimension \cite{Carone,Appelquist}, namely 'universal extra
dimension', the KK number at each vertex is conserved and, in the
present processes, the additional 'lepton-KK lepton-KK Higgs '
vertices appear. In our calculations we take into account such
vertices and assume that the Yukawa couplings existing are the
same as the ones existing in the zero-mode case. Since the extra
dimension is compactified on a an orbifold such that the zero mode
leptons and Higgs fields  are 4-dimensional model III particles,
there exist a compactification scale $1/R$, where $R$ is the size
of the extra dimension. This parameter needs to be restricted and
the lower bound for inverse of the compactification radius is
estimated as $\sim 300\, GeV$ \cite{Appelquist}.

In our work we predict the one UED effect on the BR of the LFV
processes $\mu\rightarrow e \gamma$ and $\tau \rightarrow
\mu\gamma$ in the framework of the type III 2HDM. These decays
exit at least in the one loop level in the model III and the
addition of one spatial dimension results in new loop diagrams
induced by the KK excitations of the lepton and Higgs fields (see
Fig. \ref{fig1}). In the numerical calculations we study the BR
and the ratio of KK mode contributions to the zero mode ones,
within a wide range of the compactification scale $1/R$ and try to
estimate the effects of possible complexity of Yukawa couplings
and the mass ratio of neutral Higgs bosons, $h^0$ and $A^0$.

In Fig. \ref{BRmuegamR}, we present the compactification scale
$1/R$ dependence of the BR for the  LFV decay $\mu\rightarrow e
\gamma$ for $m_{h^0}=85\, GeV$, $m_{A^0}=95\, GeV$,
$\bar{\xi}^{D}_{N,\tau\mu}=30\, GeV$, for four different values of
the coupling $\bar{\xi}^{D}_{N,\tau e}$. The solid (dashed, small
dashed, dotted) line represents the BR for $\bar{\xi}^{D}_{N,\tau
e}=0.5\times 10^{-3}(1.0\times 10^{-3},\,0.5\times 10^{-2},
\,1.0\times 10^{-2})\,GeV$. This figure shows that the BR is not
sensitive to the scale $1/R$ and therefore the contribution due to
the one spatial extra dimension is suppressed. In the BR the main
contribution comes from the term which is proportional to the one
including the function $F$ (see eq. {\ref{A1A2}). However, the
term including the extra dimension contribution is proportional
the suppression factor $x_{\mu\tau}=\frac{m_\mu}{m_\tau}$. In
addition to this the large compactification scale $1/R$ causes to
decrease the extra dimension contribution.

Fig. \ref{BRtaumugamR} is devoted to the compactification scale
$1/R$ dependence of the BR for the  LFV decay $\tau\rightarrow \mu
\gamma$, for $m_{h^0}=85\, GeV$, $m_{A^0}=95\, GeV$,
$\bar{\xi}^{D}_{N,\tau\mu}=30\, GeV$, for six different values of
the coupling $\bar{\xi}^{D}_{N,\tau \tau}$. The solid (dashed,
small dashed, dotted, dot-dashed, double dotted) line represents
the BR for $\bar{\xi}^{D}_{N,\tau \tau}=50\,(100,\, 150,\, 200,\,
250,\, 300)\,GeV$. In this figure, it is shown that the BR is not
sensitive to the scale $1/R$ for its large values. In the BR of
this decay, the term including the extra dimension contribution
does not have the suppression factor $x_{\mu\tau}$ and therefore
the sensitivity of the BR is greater for small values of the
compactification scale, compared to the one for the
$\mu\rightarrow e \gamma$ decay. However, at the large scale $1/R$
the extra dimension contribution is negligible.

In Fig. \ref{BRtaumugamksitautauR}, we present the relative
behaviors of the coupling $\bar{\xi}^{D}_{N,\tau\tau}$ and the
compactification scale $1/R$ for the fixed values of the upper
limits BR of the decay $\tau\rightarrow \mu \gamma$, for
$m_{h^0}=85\, GeV$, $m_{A^0}=95\, GeV$,
$\bar{\xi}^{D}_{N,\tau\mu}=30\, GeV$. The solid (dashed, small
dashed) line represents the upper limit of the BR as
$10^{-6}\,(10^{-7},\,10^{-8})$. It is observed that the increasing
values of the the scale $1/R$ forces the Yukawa coupling to
increase to be able to reach the numerical value of the upper
limit for the BR of the process and the sensitivity to the scale
$1/R$ increases with the increasing values of the upper limit of
the BR.

At this stage we take the Yukawa coupling
$\bar{\xi}^{D}_{N,\tau\tau}$ complex (see eq. (\ref{xi})) and
study the effects of the complexity of this coupling and the mass
ratio of neutral Higgs bosons, $h^0$ and $A^0$, to the ratio of KK
mode contributions to the zero mode ones,
$Ratio=\frac{BR^{extr}}{BR^{2HDM}}$.

Fig. \ref{BRtaumugamRatiosintet} represents the parameter
$sin\theta_{\tau\tau}$ dependence of the ratio $Ratio$ where
$BR^{extr}$ ($BR^{2HDM}$) is the contribution of the extra
dimension effects (the contribution without the extra dimensions),
for $m_{A^0}=95\, GeV$ and for five different values of the scale
$1/R$. The solid (dashed, small dashed, dotted,dot-dashed) line
represents the $Ratio$ for $1/R=100\,(200,\, 300,\,400,\,500)\,
GeV$. This ratio is of the order of $10^{-4}-10^{-3}$ for the
intermediate values of the scale $1/R$ and slightly increases with
the increasing complexity of the Yukawa coupling
$\bar{\xi}^{D}_{N,\tau\tau}$.

Finally, in Fig. \ref{BRtaumugamRatioR0} (\ref{BRtaumugamRatioR05}
and \ref{BRtaumugamRatioR1}) we present the compactification scale
$1/R$ dependence of the $Ratio$ for $sin\theta_{\tau\tau}=0$
($sin\theta_{\tau\tau}=0.5$, $sin\theta_{\tau\tau}=1$) and for
$m_{A^0}=95\, GeV$. The solid (dashed, small dashed) line
represents the dependence for the mass ratio
$r=\frac{m_{h^0}}{m_{A^0}}=0.8 \, (0.9, 0.95)$. These figures show
that the $Ratio$ increases with the increasing mass degeneracy of
neutral Higgs bosons $h^0$ and $A^0$, especially for the
increasing values of the complexity of the Yukawa coupling. This
is an interesting result since the contribution comes from the
extra dimensions are sensitive to the mass ratio of neutral Higgs
bosons $h^0$ and $A^0$ and also the complexity of the Yukawa
coupling. Notice that the extra dimension contributions reach
almost to the ordinary ones if the masses of $h^0$ and $A^0$ are
nearly degenerate and the complexity of the coupling is high.

Now we would like to present the results briefly.
\begin{itemize}
\item The BR of $\mu\rightarrow e \gamma$ decay is not sensitive
to the scale $1/R$ since the term including the extra dimension
contribution is proportional the suppression factor
$x_{\mu\tau}=\frac{m_\mu}{m_\tau}$ and the large compactification
scale $1/R$ causes to decrease the extra dimension contribution.
The BR of $\tau\rightarrow \mu \gamma$ decay is not sensitive to
the scale $1/R$ for its large values. In the BR of this decay, the
term including the extra dimension contribution does not have the
suppression factor $x_{\mu\tau}$ and therefore the sensitivity of
the BR is stronger for small values of the compactification scale,
compared to the one of the $\mu\rightarrow e \gamma$ decay.
\item The contribution comes from the extra dimensions is
sensitive to the mass ratio of neutral Higgs bosons $h^0$ and
$A^0$ and also the complexity of the Yukawa coupling.
\end{itemize}

Finally, it is not easy to investigate the extra dimension effects
in the LFV decays $BR(\mu\rightarrow e\gamma)$ and
$BR(\tau\rightarrow \mu\gamma)$, however, the more accurate future
experimental results of these decays, hopefully, will be helpful
in the determination of the signals coming from the extra
dimensions.
\section{Acknowledgement}
This work has been supported by the Turkish Academy of Sciences in
the framework of the Young Scientist Award Program.
(EOI-TUBA-GEBIP/2001-1-8)
\newpage
\begin{figure}[htb]
\vskip 2.0truein \centering \epsfxsize=6.8in
\leavevmode\epsffile{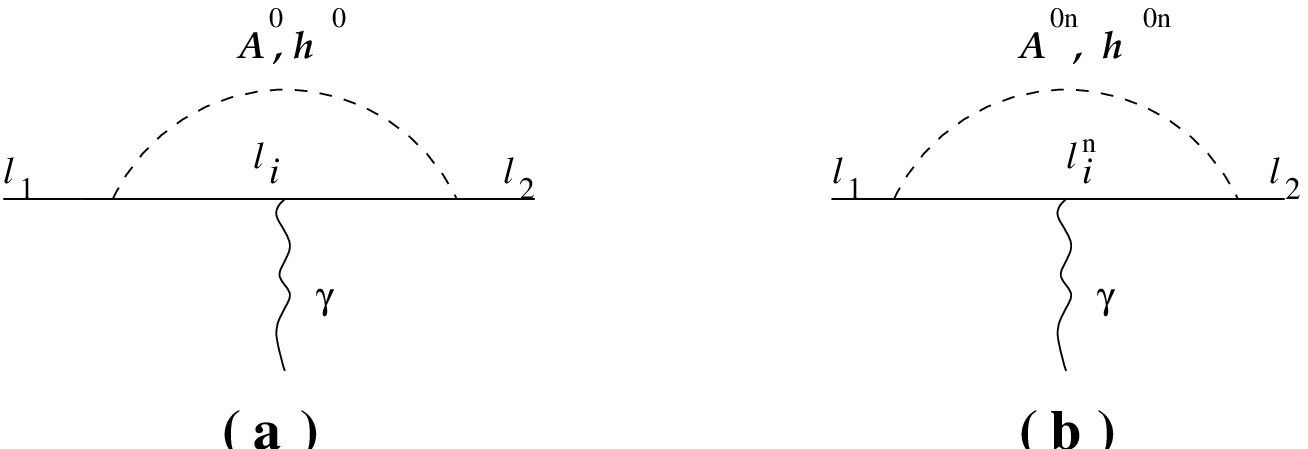} \vskip 1.0truein \caption[]{One
loop diagrams contribute to $l_1\rightarrow l_2 \gamma$ decay  due
to the zero mode (KK mode) neutral Higgs bosons $h^0$ and $A^0$
($h^{0 n}$ and $A^{0 n}$) in the 2HDM.} \label{fig1}
\end{figure}
\newpage
\begin{figure}[htb]
\vskip -3.0truein \centering \epsfxsize=6.8in
\leavevmode\epsffile{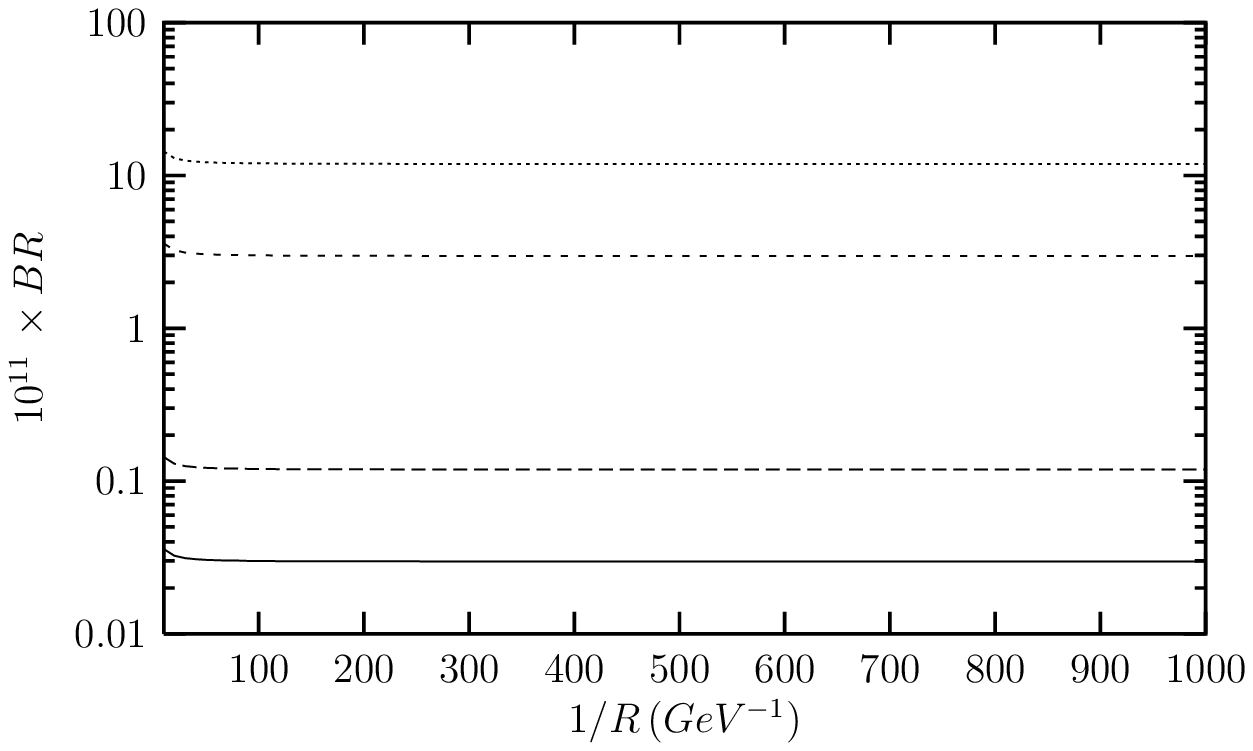} \vskip -3.0truein
\caption[]{The compactification scale $1/R$ dependence of the BR
for the  LFV decay $\mu\rightarrow e \gamma$ for $m_{h^0}=85\,
GeV$, $m_{A^0}=95\, GeV$, $\bar{\xi}^{D}_{N,\tau\mu}=30\, GeV$,
for four different values of the coupling $\bar{\xi}^{D}_{N,\tau
e}$. The solid (dashed, small dashed, dotted) line represents the
BR for $\bar{\xi}^{D}_{N,\tau e}=0.5\times 10^{-3}(1.0\times
10^{-3},\,0.5\times 10^{-2}, \,1.0\times 10^{-2})\,GeV$.}
\label{BRmuegamR}
\end{figure}
\begin{figure}[htb]
\vskip -3.0truein \centering \epsfxsize=6.8in
\leavevmode\epsffile{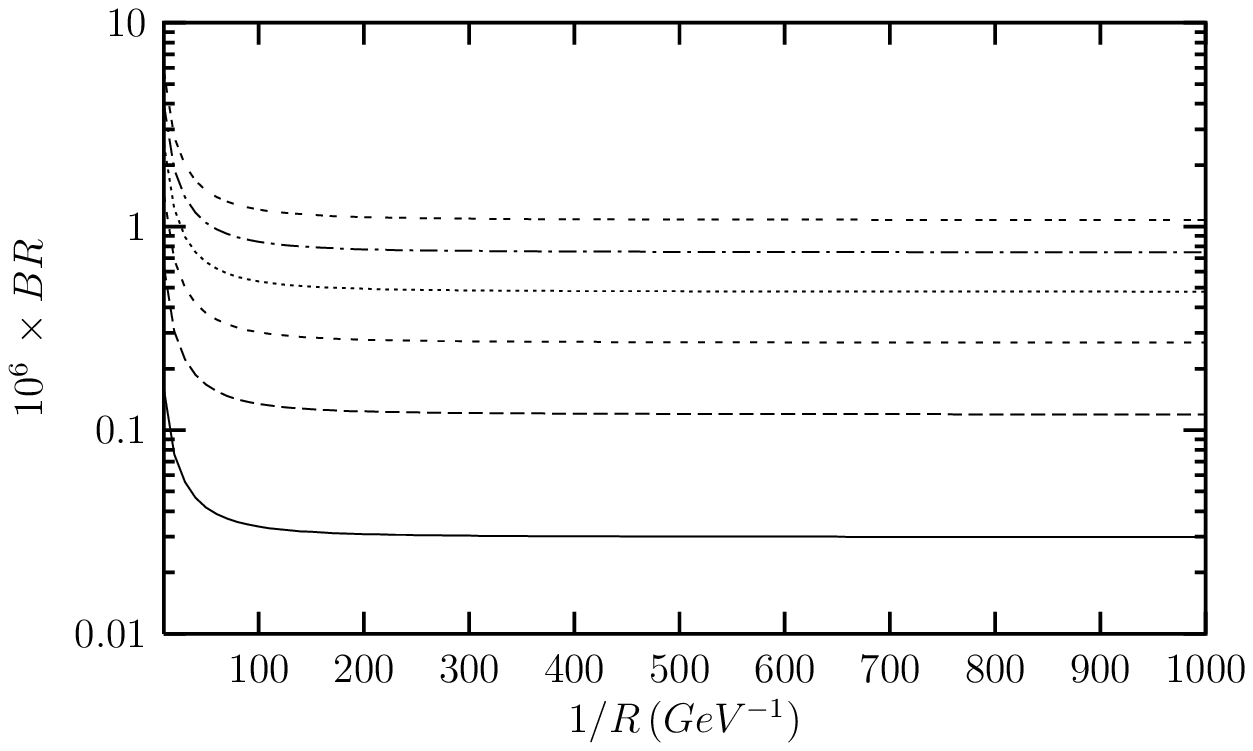} \vskip -3.0truein
\caption[]{The compactification scale $1/R$ dependence of the BR
for the  LFV decay $\tau\rightarrow \mu \gamma$ for $m_{h^0}=85\,
GeV$, $m_{A^0}=95\, GeV$, $\bar{\xi}^{D}_{N,\tau\mu}=30\, GeV$,
for six different values of the coupling $\bar{\xi}^{D}_{N,\tau
\tau}$. The solid (dashed, small dashed, dotted, dot-dashed,
double dotted) line represents the BR for $\bar{\xi}^{D}_{N,\tau
\tau}=50\,(100, \,150,\, 200,\, 250,\, 300)\,GeV$.}
\label{BRtaumugamR}
\end{figure}
\begin{figure}[htb]
\vskip -3.0truein \centering \epsfxsize=6.8in
\leavevmode\epsffile{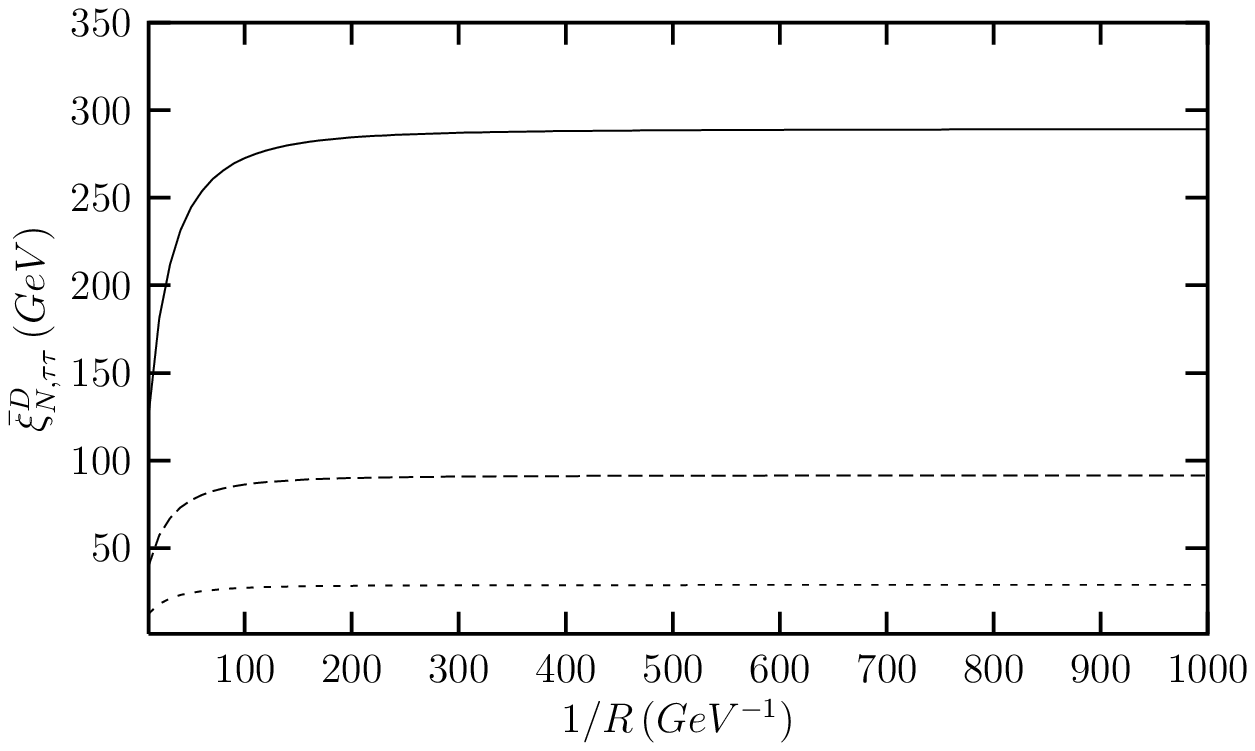} \vskip -3.0truein
\caption[]{The relative behaviors of the coupling
$\bar{\xi}^{D}_{N,\tau\tau}$ and the compactification scale $1/R$
for the fixed values of the upper limits BR of the decay
$\tau\rightarrow \mu \gamma$, for $m_{h^0}=85\, GeV$,
$m_{A^0}=95\, GeV$, $\bar{\xi}^{D}_{N,\tau\mu}=30\, GeV$. The
solid (dashed, small dashed) line represents the upper limit of
the BR as $10^{-6}\,(10^{-7},\,10^{-8})$}.
\label{BRtaumugamksitautauR}
\end{figure}
\begin{figure}[htb]
\vskip -3.0truein \centering \epsfxsize=6.8in
\leavevmode\epsffile{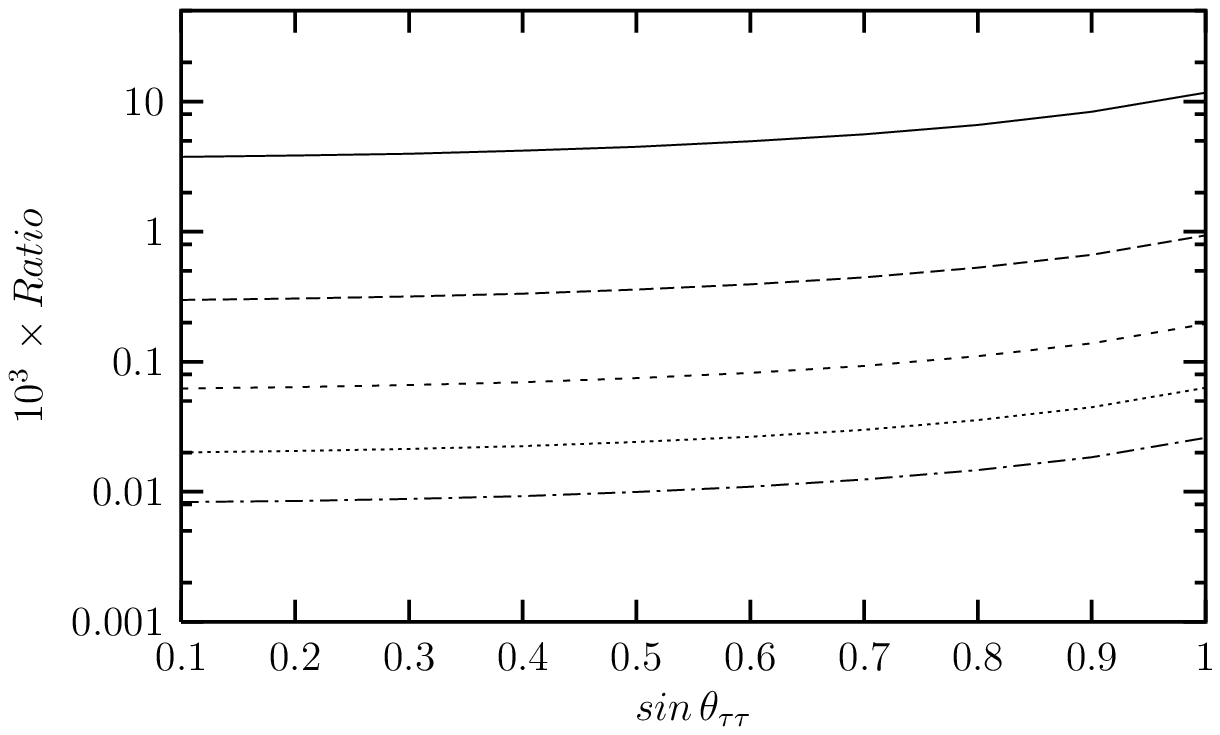} \vskip -3.0truein
\caption[]{$sin\theta_{\tau\tau}$ dependence of the
$Ratio=BR^{extr}/BR^{2HDM}$, for $m_{A^0}=95\, GeV$, for five
different values of the scale $1/R$. The solid (dashed, small
dashed, dotted,dot-dashed) line represents the $Ratio$ for
$1/R=100\,(200,\, 300,\,400,\,500)\, GeV$.}
\label{BRtaumugamRatiosintet}
\end{figure}
\begin{figure}[htb]
\vskip -3.0truein \centering \epsfxsize=6.8in
\leavevmode\epsffile{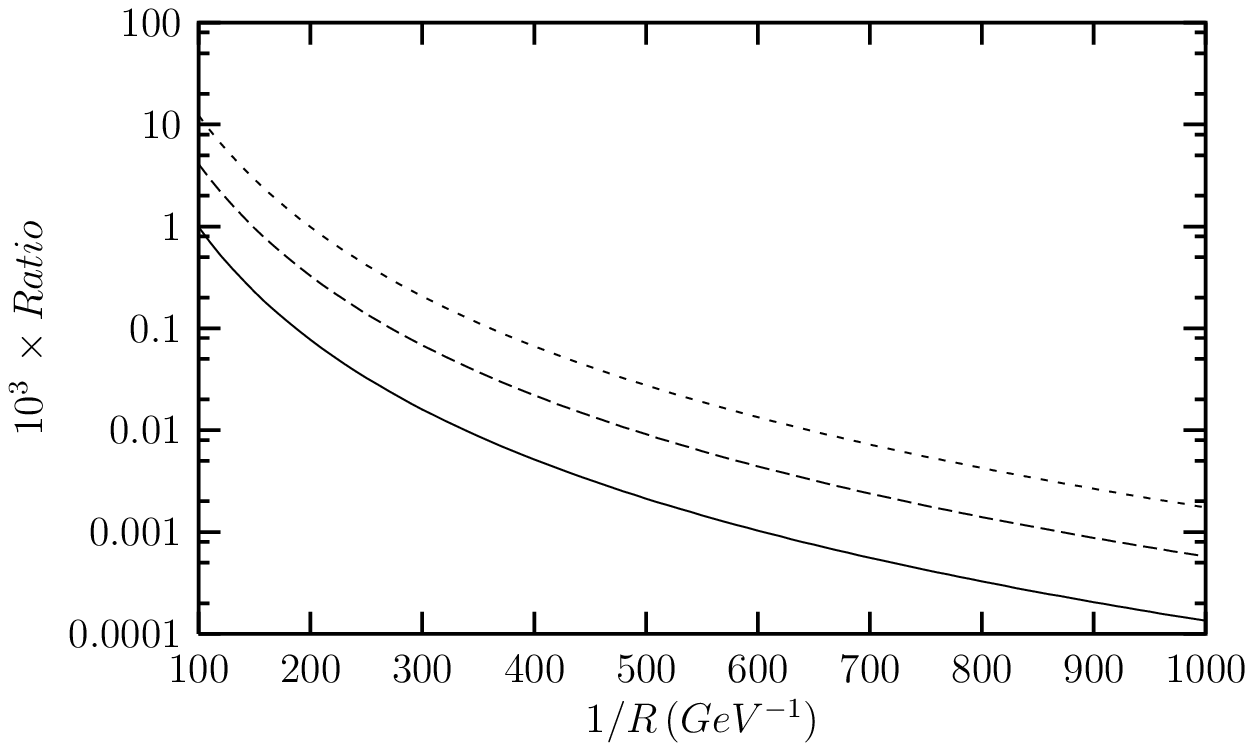} \vskip -3.0truein
\caption[]{The compactification scale $1/R$ dependence of the
Ratio for $sin\theta_{\tau\tau}=0$  and $m_{A^0}=95\, GeV$. The
solid (dashed, small dashed) line represents the dependence for
the mass ratio $r=\frac{m_{h^0}}{m_{A^0}}=0.8 \, (0.9, 0.95)$. }
\label{BRtaumugamRatioR0}
\end{figure}
\begin{figure}[htb]
\vskip -3.0truein \centering \epsfxsize=6.8in
\leavevmode\epsffile{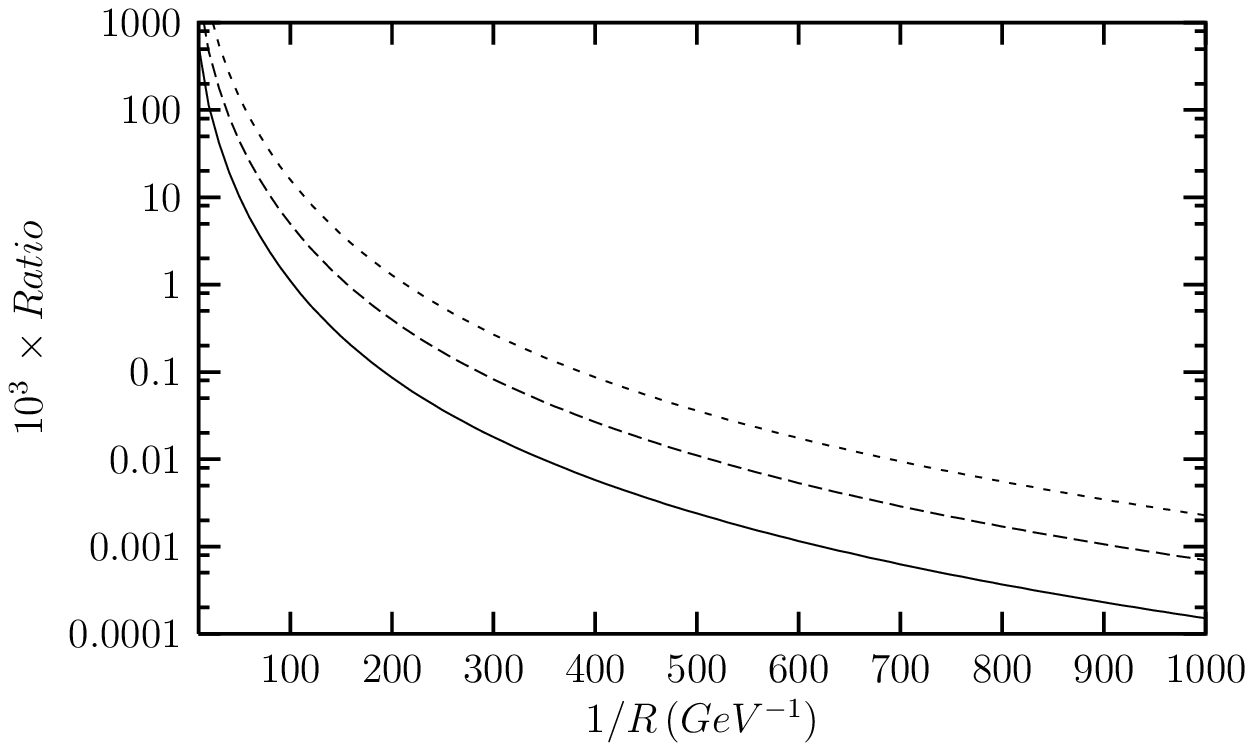} \vskip -3.0truein
\caption[]{The same as Fig. \ref{BRtaumugamRatioR0} but for
$sin\theta_{\tau\tau}=0.5$ }
\label{BRtaumugamRatioR05}
\end{figure}

\begin{figure}[htb]
\vskip -3.0truein \centering \epsfxsize=6.8in
\leavevmode\epsffile{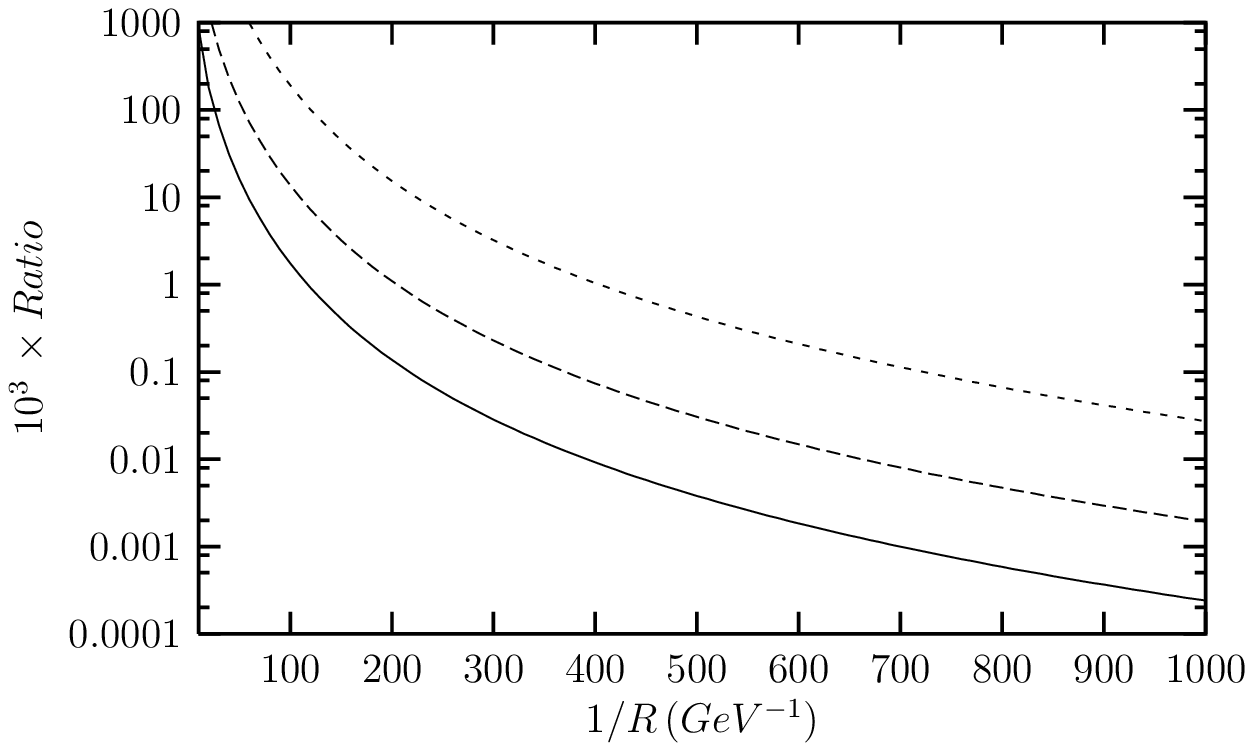} \vskip -3.0truein
\caption[]{The same as Fig. \ref{BRtaumugamRatioR0} but for
$sin\theta_{\tau\tau}=1$}
\label{BRtaumugamRatioR1}
\end{figure}
\end{document}